\documentclass{iaus}
\usepackage{graphicx}

\def\solm{M$_{\odot}\,$}

\def\solm{M$_{\odot}\,$}

\def\casgm20{CAS-G-M$_{20}\,$}
\def\m20{M$_{20}\,$}

\title[How do galaxies get their baryons?] 
{How do galaxies get their baryons?}

\author[Christopher J. Conselice]   
{Christopher J. Conselice$^1$}

\affiliation{$^1$Centre for Astronomy and Particle Theory,  Nottingham University \\ Nottingham, UK \\ email: {\tt conselice@nottingham.ac.uk} \\[\affilskip]}

\pubyear{2011}
\volume{xxx}  
\pagerange{119--126}
\jname{Tracing the Origin of Galaxies in the Land of our Ancestors}
\editors{Carignan, C., Freeman, K.C., Combes, F. eds.}
\begin{document}

\maketitle

\begin{abstract}

Understanding how galaxies obtain baryons, their stars and gas, over cosmic 
time is traditionally approached in two different ways - theoretically and observationally.  
In general, observational approaches to galaxy formation include measuring basic galaxy
properties, such as 
luminosities, stellar masses, rotation speeds, star formation rates and how these features 
evolve through time. Theoretically, cosmologically based models collate the physical 
effects driving galaxy assembly - mergers of galaxies, accretion of gas, star formation, 
and feedback, amongst others, to form predictions which are matched to galaxy observables.  
An alternative approach is to examine directly, in an observational way, the processes 
driving galaxy assembly, including the effects of feedback. This is a new `third way' towards 
understanding how galaxies are forming from gas accretion and mergers, and directly probes these 
effects instead of relying on simulations designed to reproduce observations.  
This empirical approach towards understanding galaxy formation, including the acquisition
history  of
baryons, displays some significant 
differences with the latest galaxy formation models, in addition to directly demonstrating 
the mechanisms by which galaxies form most of their baryonic mass.

\end{abstract}

\firstsection 
\section{Introduction}

Galaxies are the astronomical gift that keeps on giving.  Galaxies are used 
as locators of mass in the universe, where some of the best evidence for dark matter exists,
and are used as tracers of distance in cosmological studies of the universe
as a whole, providing us with the best evidence for dark energy.  
Galaxies themselves are however endlessly fascinating in
themselves, as they are the basic structure of the universe.  
By understanding how and when galaxies formed and evolved we
are essentially finding out how the universe itself is structured.

Many vigorous ongoing  
observational and theoretical programs are designed to answer the question of
how galaxies form and evolve.  Gravitational collapse is the obvious answer
to this question, yet we know both theoretically and observationally that 
galaxies could not have the same total or stellar mass in the early universe as
they do today.  A 
collection of processes from mergers, gas accretion, and feedback from stars
and AGN have all played a role in creating the universe of galaxies that
we see today.  We are however just starting to unravel the role of these various
processes and how they work together.

Understanding galaxy formation in a physical way has traditionally been 
approached in a theoretical manner. The first attempts to understand galaxy 
formation, such as the collapse
models of Eggen et al. (1962) have been based on observations of galaxy 
properties, such as the ages and velocities of stars in our own galaxy, to more
modern attempts using quantities such as luminosity/mass functions in the
local universe.  However, all attempts to model galaxies
with basic physics still cannot reproduce all known observations (e.g.,
Guo et al. 2010).

Contemporary approaches towards understanding galaxy formation in a 
theoretical and cosmological context are dominated by semi-analytical models 
which typically use local galaxy properties to calibrate their simulation 
output.  The parameters adjusted in
these simulations are many, and are based on initial conditions of baryons that
are still unknown.  Despite the immense effort over more than 30 years, these
simulations still have significant problems in reproducing high redshift
observations, particular for the most massive galaxies (e.g., Conselice et al.
2007; Bertone \& Conselice 2009; Guo et al. 2010).

An alternative approach is to try to understand the various physical mechanisms
that drive the formation of galaxies at high redshift, and to examine distant
galaxies to directly trace these assembly processes on the evolution of the
baryonic content of these systems.  This is an empirical approach to galaxy
evolution/formation, and its success requires that we are: 
(1) able to determine  observationally the physical process driving evolution; and 
(2) be confident that we have identified all mechanisms in which galaxies 
obtain their baryons.

In this review, I will discuss the `third way' of solving the problem
of galaxy evolution in terms of their baryons. This method involves searching for 
physical processes
of galaxy formation at high redshifts in galaxies as a function of their stellar
mass, and determining the contribution of these processes to the further
formation of the masses of galaxies at later times. By 
continuing this processes at all redshifts, where observations can reveal 
ongoing physical processes,  we are measuring how galaxies are forming in a 
statistical sense as a function of galaxy mass.  

For this `third way' approach to work, it is necessary to be certain that
we have identified the major processes by which galaxies form, and to have
some confidence that we can identify these processes through 
observations.  The first is easier to show -- we are fairly certain that the
dominant methods by which galaxies form are now identifiable.  However, what is
certainly not understood is the relative contributions of these various processes
towards forming galaxies.

The processes that we believe to be dominant in settling baryons into galaxies
and converting them into stars include: major and minor mergers, gas accretion,
and gas cooling of residual gas from the initial collapse of
the baryons and its dark matter halo.  On the other hand, we now know that 
galaxy feedback, in the form of energy ejection, is a major process in 
regulating galaxy formation and evolution. This feedback can
come in many forms including, and especially, feedback from stars, most
prominently in the form of supernova, and feedback from energy emitted by
a central massive AGN.  Each of these processes in principle
has an observational signature.  In some cases, such as cooling of residual
gas and gas accretion, direct signatures are difficult to find, however 
circumstantial evidence for these are obtainable.

Perhaps the most straightforward method that can be traced are major
merger signatures, which can now be found using both galaxy pair and morphological
methods (e.g,. Conselice et al. 2003, 2008) and through the kinematics of gas
in these systems (e.g., Genzel et al. 2008).  In general, it has been shown for
typical galaxies of stellar masses 10$^{9-10}$ \solm that the total number of
major mergers is roughly $\sim 4$ at $z < 3$ (Conselice et al. 2008).  However,
this amount of merging is not enough to account for the increase in the
masses of galaxies during this epoch (Mortlock et al. 2011), and other 
formation processes are necessary.  

\begin{figure}[b]
\begin{center}
 \includegraphics[width=4in]{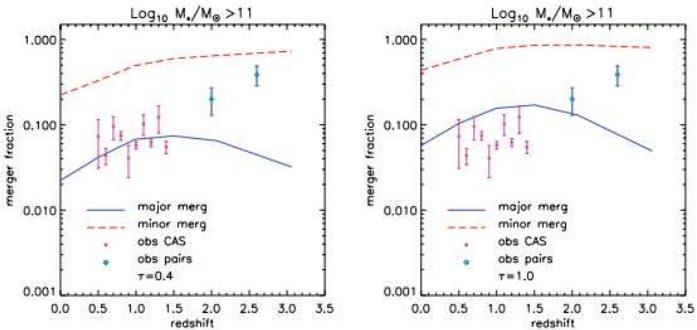} 
 \caption{Figure showing the merger history for massive galaxies with
stellar masses, M$_{\odot} > 10^{11}$ \solm, located at $z < 3$.  The
points show measurements from different surveys, and the increase
in the merger fraction goes as $\sim (1+z)^{3}$.  The long dashed line shows
the predicted major merger history from the Millennium simulation 
(see Bertone \& Conselice 2009). The right and left panel show the
range of predicted vs. observed merger fractions using different time-scales ($\tau$).}
   \label{fig1}
\end{center}
\end{figure}

Within this proceeding I will describe attempts to construct the
baryonic assembly of galaxies using new data sets and programs focused
on very massive galaxies with M$_{\odot} > 10^{11}$ \solm at $z < 3$, including
the {\em GOODS NICMOS Survey} (GNS).

\section{Evolution of Massive Galaxies from the GOODS NICMOS Survey}

\begin{figure}[b]
\begin{center}
 \includegraphics[width=4.7in]{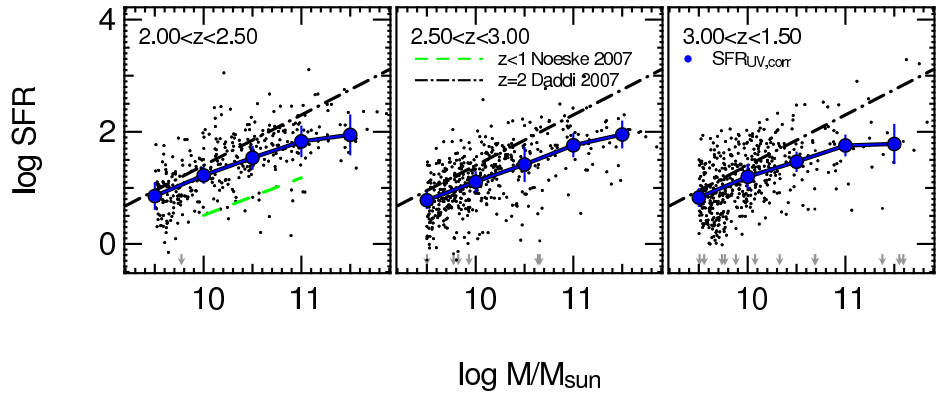} 
 \caption{The star formation rate as a function of stellar mass from $z = 1.5$
to $z = 3$ using dust corrected ultraviolet light. The blue line shows the average 
values for GNS points,
while comparisons to previous work is also show (see Bauer et al. 2011). }
   \label{fig1}
\end{center}
\end{figure}

Much of the data discussed henceforth originates from papers within the
GOODS NICMOS Survey (GNS). This is a near-infrared survey of massive
galaxies selected by M$_{\odot} > 10^{11}$ \solm located at redshifts
$1.5 < z < 3$.  The  depth of these images is $H_{\rm AB} = 26.8$ (5 $\sigma$), and the
resolution is high enough to make basic morphological measurements, such as
sizes and concentration (Buitrago et al. 2008; Conselice et al. 2010).
This survey, combined with a lower redshift sample of similar mass galaxies
from the POWIR/DEEP2 Survey (Conselice et al. 2007; Trujillo
et al. 2007) allows us to trace galaxy formation processes for galaxies
with masses  M$_{\odot} > 10^{11}$ \solm from $0.4 < z < 3$.  

We primarily
examine massive galaxies for two reasons. The first is that these systems
are the easiest to study since they are often the brightest. The second reason
is that predictions of galaxy formation models are often most reliable for
the most massive systems.  In the lambda CDM model these are also the galaxies
which are most evolved, and are therefore outstanding test-beds for
comparing observations to theory.  We discuss below the observations we
use to derive the baryonic history of these massive galaxies, and later we combine
this information into a summary for how very massive galaxies have formed
and developed throughout time.  In this review we summarize findings on the merger
history of these galaxies, the AGN and star formation history, as well as circumstantial
evidence of gas accretion.

\begin{figure}[b]
\begin{center}
 \includegraphics[width=5.6in]{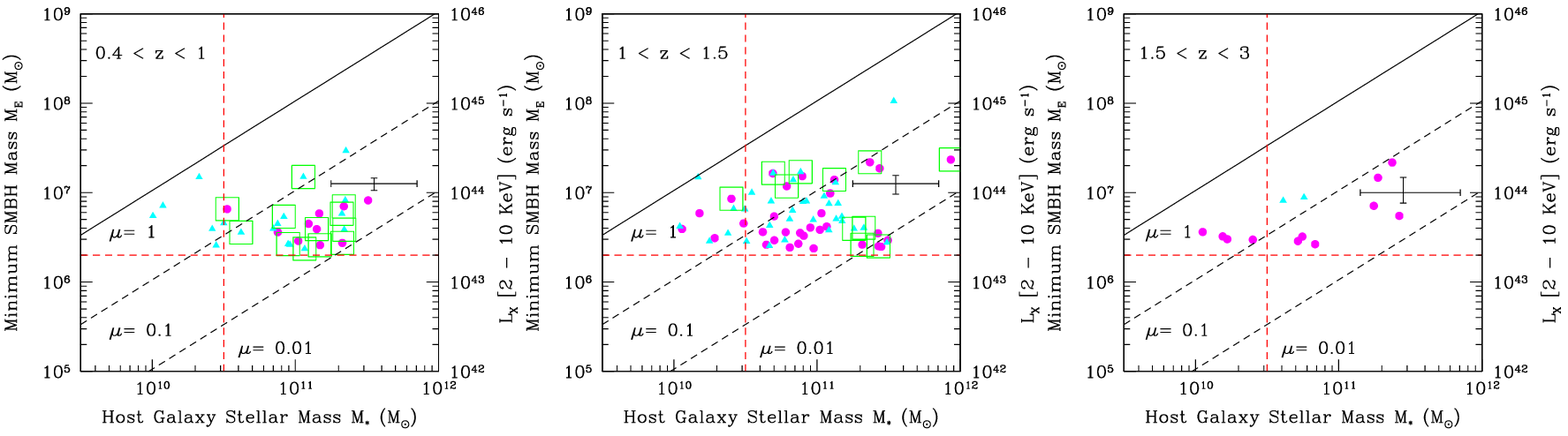} 
 \caption{Plots of the evolution of the minimum black hole mass ($M_{\rm E} = \mu M_{\rm BH}$),
based on X-ray luminosities as a function of galaxy stellar mass for a galaxy selected 
sample up to $z = 3$. Shown are lines for the relation between black hole and galaxy mass 
at different values of $\mu$ (Haring \& Rix 2004).  Magenta circles are 
hard X-ray sources and cyan triangles are soft sources with those surrounded
by green boxes systems with spectroscopic redshifts (Bluck et al. 2011).}
   \label{fig1}
\end{center}
\end{figure}

\subsection{The Galaxy Merger History}

The major merger history of a sample of very massive galaxies is described in
Bluck et al. (2009) who find that while the major merger fraction for
M$_{\odot} > 10^{11}$ \solm galaxies at $z < 1.5$ is roughly 10\%, at higher redshifts,
this increases to 20\%-30\% at $z = 2-3$
(Bluck et al. 2009; Conselice et al. 2009; see Fig. 1).  This is roughly consistent
with other observations of the merger history using a variety of methods (e.g.,
Conselice et al. 2008).    However, to derive the
role of mergers in galaxy formation, it is necessary to calculate the
merger rate, that is the number of mergers occurring per unit time as
a function of stellar mass.  Obtaining an accurate measure of this
merger rate requires the time-scale for merging to be
known with some certainty (e.g., Bertone \& Conselice 2009).

The merger rate for massive galaxies, as derived from simulations of
both pairs and morphology, are consistent with each other (Conselice
et al. 2009).  For these very massive galaxies in
major merger pairs, the time-scale is 0.5-1 Gyr, depending on the
mass and mass ratio of the merger and the method used to find these mergers (e.g.,
Lotz et al. 2008).  Using this to calculate merger
rates, Bluck et al. (2009) find that on average a galaxy will undergo
1.7$\pm0.5$ major mergers at $z < 3$.  This will roughly double the
stellar mass within these galaxies over this time period.  The role of minor
mergers in building  up the masses of galaxies is still largely unknown,
but is likely at a similar level (Bluck et al. 2011).  Neither minor nor
major mergers however can increase the amount of mass within massive
galaxies with M$_{\odot} > 10^{11}$ \solm. However, it is possible
that these minor mergers are able to increase the sizes of these
massive systems, which are well known to be very compact at
$z > 1$ (e.g., Buitrago et al. 2008).

\subsection{Star Formation History and Gas Accretion History}

The star formation history of the universe is one of the classical ways
to study galaxy evolution.  We are now able to trace the star formation
history in many ways using many techniques, and  we now
can examine this as a function of
galaxy property and redshift. It is now well established, for example, that the ongoing 
star formation rate scales with galaxy stellar mass, such that the most
massive galaxies are undergoing the highest star formation rates. 
Systems with M$_{\odot} > 10^{11}$ \solm have star formation rates 
between 100 - 500 \solm year$^{-1}$
on average (Daddi et al. 2007; Bauer et al. 2011).  
What is interesting 
about these observations is that the
star formation rate for M$_{\odot} > 10^{11}$ \solm systems 
has a low scatter for those systems which are star forming.
That is, either these massive galaxies are not undergoing star formation
at all, or they contain a very high star formation rate, which indicates that the
cold gas mass of these systems scale with their stellar mass and galaxy
wide star formation events are occurring in tandem.

The increase of stellar mass from star formation can be calculated by integrating the
star formation rate for these massive galaxies over time.  Doing this, we
find that the stellar mass added to galaxies through star formation will
roughly double the stellar mass of these systems from $1.5 < z < 3$.  This is
potentially a very fundamental observation, as it tells us that some process is required
to not only keep the star formation active, but also that enough gas is
reacquired by galaxies to become replenished in some way for this star formation
to continue.

What triggers this star formation is still uncertain in some
individual galaxy cases, but we can determine 
a few likely properties of the star formation and where it originates based
on the likely amount of cold gas within these galaxies as a function of
time, and the observation that the star formation rate for M$_{\odot} > 10^{11}$ \solm
galaxies is constant over $1.5 < z < 3$. 

At the same time, we know that the cold
gas mass density is roughly constant as well, and quite low at 10\% for 
galaxies as massive as our systems with M$_{\odot} > 10^{11}$ \solm (e.g.,
Erb et al. 2006; Daddi et al. 2010).  If the amount of cold gas available to produce
new stars is  $< 10$\% of the current stellar mass up to $z \sim 3$, then it is
very difficult, if not impossible, to produce the amount of observed star formation
in these systems without gas replenishment. Even if gas rich minor mergers are occurring 
it will still be
difficult to not only produce this much star formation, but also to keep the
cold gas supply constant.  One possible method for this is cold gas accretion from
the intergalactic medium (e.g., Dekel et al. 2009).

\subsection{Feedback from Central Black Holes}

As is well known, most galaxies have a black hole at their centers whose mass
is proportional to that of the galaxy itself (e.g., Haring \& Rix 2004). The
reason behind this strong relation is not understood completely, nor observationally
how the relation between the masses of central black holes and their host
galaxies evolve with time. However, we do know that the more massive a central
black hole is, the more accretion that black hole has had over time, and therefore
the more energy from an active galactic nuclei.  

The amount of energy ejected
from this AGN into the galaxy is unknown. However, by examining the frequency of
AGN over time, we can obtain some estimate for the amount of energy emitted
by these forming AGN into the galaxy.  As Bluck et al. (2011) show, at any
one time, 7.4$\pm$2 percent of galaxies with M$_{\odot} > 10^{10.5}$ \solm
at $z \sim 2.5$ have an X-ray selected AGN with $L_{\rm X} > 2.35 \times
10^{43}$.  By using the X-ray luminosities of these sources, we can obtain
a measure of the central black hole mass times the Eddington ratio ($\mu$) (Fig. 3).
Using a measure of these masses times Eddington ratios, we can obtain a minimum
amount of energy that these black holes have produced in the form of X-ray AGN
since $z = 3$ (Bluck et al. 2011). The result of this is that the amount of energy
ejected from the AGN integrated over time, such that the masses of these black holes
reach their value today, is roughly a factor of ten times that of the binding energy
of the galaxy. While we cannot say how this energy couples with the stars or gas in
these galaxies, this amount of energy must have a profound effect on the evolution of
these systems.

\section{Comparison with Theory and Models}

The results achieved thus far are in direct competition with 
models which predict the same quantities. We can test how our observational
results of these processes compare with semi-analytical models based on the
Millennium simulation (Bertone \& Conselice 2009).  One initial result 
of this comparison is shown in Figure~1.  This shows how the predicted
merger fraction various with redshift from the Millennium simulation compares
with various merger fraction measurements from Conselice et al. (2003),
Conselice et al. (2008, 2009) and Bluck et al. (2009).

The result of this comparison shows that the simulation results of mergers
and the data show quite different merger histories.  The measured merger fractions
are always higher than the simulation results, showing that perhaps major
mergers as measured through structure and through galaxy pairs are more
important than what is assumed in models of galaxy formation.  

This difference might be related to several effects. The first is that we known
that the number densities of massive galaxies are higher at $z > 1$
 than that predicted in simulations -- demonstrating that galaxy formation is a much
faster process that what is predicted in these models (Conselice et al. 2007).  
The other is that the stellar masses in these models could be off, particularly
for satellite galaxies which are quenched in these models, resulting in
more minor than major mergers (Bertone \&  Conselice 2009).  In the future,
more direct observations will become possible, and these can be used to 
determine galaxy formation processes largely independently of simulations.

I thank my collaborators, who have had a critical role in the research presented here. This research was 
supported by the United Kingdom's STFC and the Leverhulme Trust.


\end{document}